\documentclass[12pt]{emulateapj}
\usepackage{epsfig,graphicx}

\begin{document}
\title{On The Color Magnitude Relation of Early-type Galaxies}			
\author{Joachim Janz\altaffilmark{$\star$} \& Thorsten Lisker}
\affil{Zentrum f\"ur Astronomie der Universit\"at Heidelberg (ZAH), M\"onchhofstra{\ss}e 12-14, D-69120 Heidelberg, Germany}
\altaffiltext{$\star$}{Fellowship holder of the Gottlieb Daimler- and Karl Benz-Foundation.}
\email{jjanz@ari.uni-heidelberg.de}

\submitted{}


\begin{abstract}
In this letter we present a study of the color magnitude relation of
468 early-type galaxies in the Virgo Cluster with Sloan Digital Sky
Survey imaging data. The analysis of our homogeneous,
model-independent data set reveals that, in all colors ($u-g$, $g-r$, $g-i$, $i-z$) similarly, 
giant and dwarf early-type galaxies follow a continuous color magnitude relation (CMR)
that is best described by an S-shape. The magnitude range and quality of our data allows us to clearly confirm that the CMR in Virgo is not linear.
 Additionally, we analyze the scatter about the CMR and find that it increases in the intermediate-luminosity regime.
Nevertheless, despite this observational distinction, we conclude from
the similarly shaped CMR of  semi-analytic model predictions that dwarfs and giants
could be of the same origin.
\end{abstract}


\section{Introduction}

That early-type galaxies form a well-defined Color Magnitude Relation (CMR) 
was first recognized by \citet{1959PASP...71..106B}. Extensive studies by
\citet{1977ApJ...216..214V} and \citet{1978ApJ...223..707S,1978ApJ...225..742S}
compared the CMR of nine galaxy clusters and of field galaxies. They
found them to be astonishingly similar.
In combination with other studies comparing the CMR for E
and S0, within clusters, groups and the field
\citep{1973ApJ...179..731F,1992MNRAS.254..601B} a high degree of
universality was shown for the relation.

The CMR is typically explained by an increase of mean stellar
metallicity with increasing galaxy mass as the dominant effect.
The common underlying idea is that more massive galaxies have deeper potential
wells, which can retain metal-enriched stellar ejecta more effectively
and subsequently recycle the enriched gas into new stars
\citep{1997A&A...320...41K,1999ApJ...521...81F,2006MNRAS.370.1106G,2006MNRAS.366..717C}.
Another factor could be a variable integrated galactic initial mass function,
with more massive stars in more massive galaxies, and thus a more substantial enrichment
(e.g.~\citealt{1998A&A...335..855M,2007MNRAS.375..673K}).

An alternative explanation was given with a change of the mean
  age (see e.g. \citealt{2001ApJ...562..689P}, and some observational
  support for it in \citealt{2004AJ....127.1502R}).
The question of which explanation to favor is not totally settled yet due
  to the ambiguity introduced by the age-metallicity degeneracy. But
  data from clusters at higher redshift show that the CMR is in place
  ever since, favoring the mass-metallicity relation to be the dominant
  effect \citep{1997A&A...320...41K}.

From early on, it was
discussed whether the universality of the CMR also holds over the
whole range of galaxy masses,  
i.e.~whether dwarf and giant early-type galaxies follow the same
CMR. Studies of different clusters show consistency with one
common CMR for dwarfs and giants, albeit with a
  significant increase in the scatter at low luminosities 
(\citealt{1997PASP..109.1377S} for Coma, \citealt{2002AJ....123...2246C} for
Perseus, \citealt{2003MNRAS.344..188K} and \citealt{2007A&A...463..503M} for Fornax,
\citealt{2008MNRAS.386.2311S} for Antlia and \citealt{2008A&A...486..697M}
for Hydra I).
More explicitely, \citet{1983AJ.....88..804C} stated that there is a
common linear relation. 
But his Fig.~3 might hint at a change of 
slope from high to low luminosities, similar to what
\citet{1961ApJS....5..233D} suggested. Interestingly,
visual examination of the diagrams presented by most of the above-mentioned
studies indicates consistency also with a change of slope -- yet 
linear relations were fitted in most cases
(see, however,
\citealt{2006ApJS..164..334F}; see Sect.~5).
In this letter we revisit the question of the universality of the CMR for 
dwarfs and giants.

\begin{figure*}
\includegraphics[scale=.65,angle=-90]{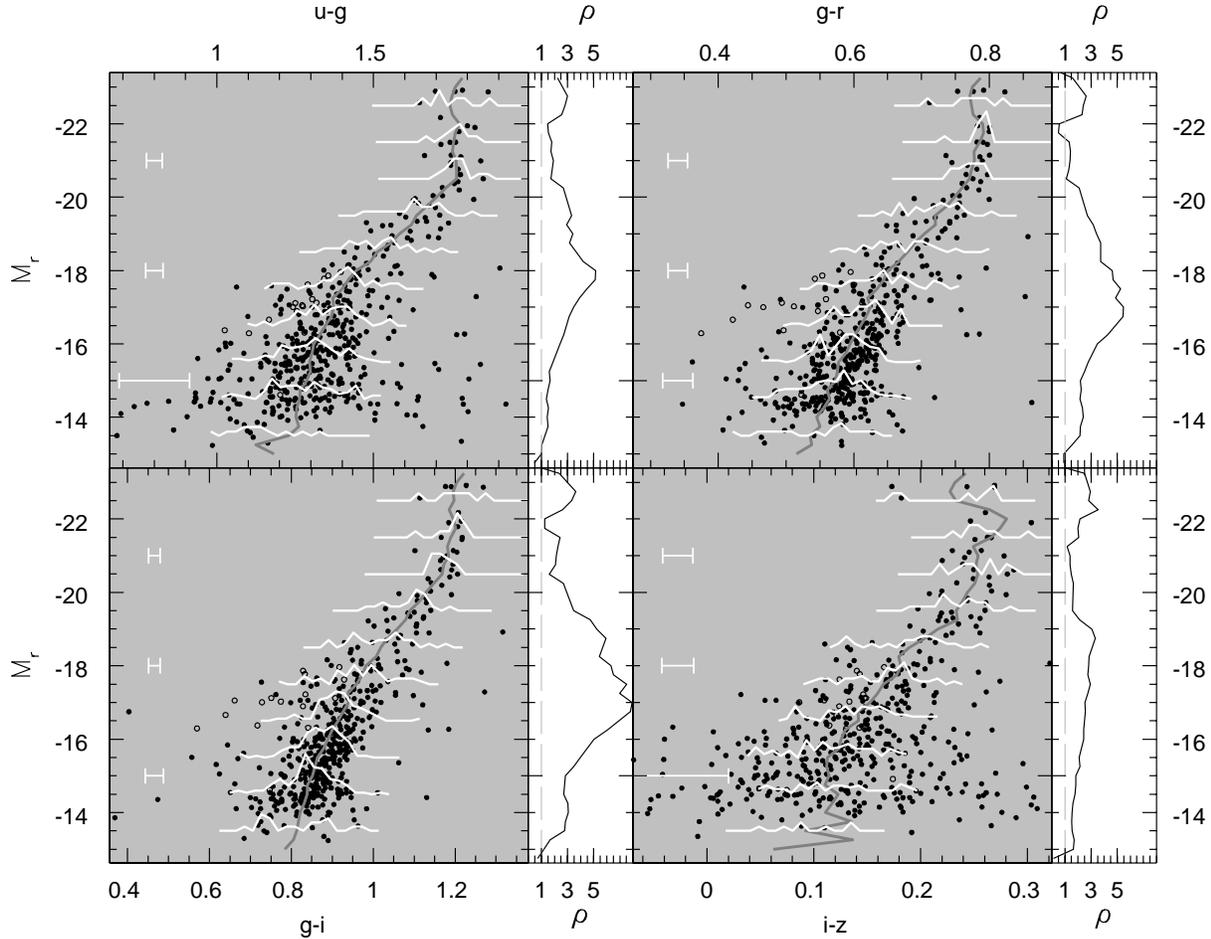}
\caption{Color Magnitudes Relations for the colors $u-g$, $g-r$,
    $g-i$, and $i-z$. \emph{Open circles} show dEs with blue cores
and \emph{filled circles} all the other galaxies in the
    sample. Limits for the x-axes are scaled to the mean color of galaxies 
fainter than $M_r>-15$ mag and brighter than $M_R<-21$ mag. 
The gray line indicates the ``running histogram'' as found
    in successive magnitude bins with a width of $1$ mag and steps of
    $0.25$ mag, clipped one time 
at $3 \sigma$.  We limit the drawing range for the
line to the region with at least three galaxies in a bin.
The white histograms show the
    distributions in bins of the same width, normalized to the square root of the number of galaxies in the bin, shown for every fourth step. 
The white errorbars indicate typical photometric errors at the respective brightness.
In the panels right
    to the color magnitude diagrams a measure for the intrinsic scatter is given by the
    ratio of the RMS of the scatter and of the photometric error in the same running bins (see
    text). }
\end{figure*}

\section{Sample Selection and Data}

The sample is based on the Virgo Cluster Catalog (VCC; \citealt{b_s_t}). 
All early-type galaxies therein with a certain cluster member status and $m_B<18.0$ mag are taken into account,
which is the same magnitude limit up to which the VCC was found to be
complete.  
This translates into $M_B<-13.09$ mag with our adopted distance modulus
of m-M=31.09 mag (d=16.5 Mpc,
\citealt{2007ApJ...655..144M}).
We treat uncertain classifications as in \citet{2008ApJ...689L..25J}
and exclude possible irregulars.
Seven further  galaxies are
  exluded due to their low $S/N$ in $u$ and $z$.

The Sloan Digital Sky Survey (SDSS)
Data Release Five (DR5) \citep{2007ApJS..172..634A} covers all but six
early-type dwarf galaxies of the VCC. 
Since the quality of sky level subtraction of the SDSS pipeline 
is insufficient, we use sky-subtracted images as provided by
\citet{lisker_etal}, based on a careful subtraction method.
The images were flux-calibrated and corrected for galactic extinction
\citep{1998ApJ...500..525S}.  
 
For each galaxy, we determined a ``Petrosian semimajor axis"  (\citealt{1976ApJ...209L...1P}, 
hereafter Petrosian SMA, $a_p$), i.e., we
use ellipses instead of circles in the calculation of the Petrosian  
radius (see, e.g., \citealt{2004AJ....128..163L}). The total flux in the
$r$-band was measured within $a = 2 a_p$, yielding a value for 
the half-light semimajor axis, $a_{hl,r,uncorr}$. 
This Petrosian aperture still misses some flux, which is of particular
relevance for the giant galaxies \citep{2001MNRAS.326..869T}. The
brightness and the half-light SMA were
corrected for this missing flux according to
\citet{2005AJ....130.1535G}.\footnote{The correction is computed with the assumption and fitting of S\'ersic profiles as in 
\citet{2008ApJ...689L..25J}. But the effect on the CMR is small: for a de Vaucouleurs profile it leads to a difference of 0.2 mag in brightness and practically no difference in color, since the correction to the radius has a minute effect due to the small color gradients.}
 Axial ratio and position angle were then
determined through an isophotal fit at  $a = 2 a_{hl,r}$.
Colors were measured within the elliptical $r$-band half-light aperture for each filter. 
Errors were estimated from the $S/N$ and calibration uncertainties (which we estimate to have a \emph{relative} effect of 0.1 mag in each band, which is smaller than the absolute values given by SDSS), as described in \citet{2008AJ....135..380L}.

Our data constitute a very homogeneous set of measurements for galaxies in one cluster, 
obtained and reduced with the same instrument, setup, and software.

\section{The CMR of early-type galaxies}
 
From the five SDSS filter bands we choose four
representative colors: $u-g$, mostly sensitive to age; $g-r$ with
the highest $S/N$; $g-i$ with the longest wavelength baseline at
good $S/N$; and $i-z$, mostly
sensitive to metallicity.

First of all, the impression one can get by examining just the \emph{black
  points} in Fig.~1 is that there is not one common linear relation from the faint to the
bright galaxies. In all colors the overall shape appears more like
  ``S'' shaped.
The brightest ($M_r<-21$) 
galaxies have almost constant color, i.e.~no correlation between color and brightness; the very brightest galaxies
show a larger scatter. These were reported before to be \emph{morphologically}
different from the other galaxies in more detailed studies of the
inner light profiles
(e.g. \citealt{2001AJ....122..653R,2004AJ....127.1917T,2006ApJS..164..334F,2007ApJ...664..226L,2008arXiv0810.1681K}). For the remaining galaxies
several descriptions seem to be plausible, ranging from just an offset
  between two relations with similar slopes up to a curved relation.

\begin{figure}
\includegraphics[scale=.3,angle=-90]{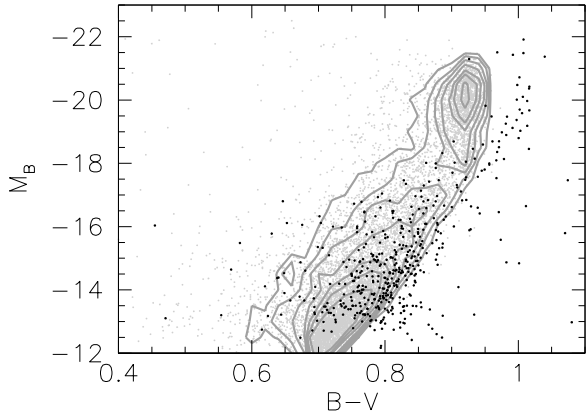}
\caption{Color Magnitude Diagram, comparison to the semi-analytic model. Virgo cluster galaxies are shown with black symbols and model
  galaxies from the Numerical Galaxy Catalog
  \citep{2005ApJ...634...26N} as gray dots (we selected galaxies with $B/T > 0.6$ and with higher effective surface brightness than 25.5 mag arcsec$^{-2}$). Contours were
  calculated using model galaxy abundances in bins of 0.5 mag in brightness and 0.02 
  mag in color. The contour levels, relative to the lowest level, are
  2, 3.4, 5, 7.5 10, and 15. }
\end{figure}

With the non-linear shape,
it seems not very favorable  to fit a straight line. This would not
describe the data well, and 
there is no theoretical prediction what other function is expected. So
at first, we want to make the overall shape more clearly visible, using
continous, overlapping magnitude bins, in which mean and scatter are calculated.
In Fig.~1 these derived relations are shown with grey lines. The first impression is
confirmed: one common linear relation for dwarfs and giants cannot be
seen. Moreover, the white histograms showing the galaxy distributions in 
the bins are clearly peaked towards the bright and the faint end,
while they are rather flat at intermediate luminosity.

If the notion of two separate relations with an overlap is correct,
this should become evident through an increased scatter about the
relation around the intersection, relative to the photometric error. 
Therefore we calculate the RMS of the scatter around the mean in
running bins (clipping one time at 3$\sigma$) and divide it by the RMS of the photometric errors 
$$\rho\equiv\textrm{RMS-ratio}\equiv\frac{\textrm{rms}_\textrm{scat}}{\textrm{rms}_\textrm{err}}=\frac{\sqrt{\sum_i
    (c-\left<c\right>)^2}}{\sqrt{\sum_i \sigma_i^2}},$$
with color $c$ and mean color $\left<c\right>$, 
averaging over the galaxies in the respective bin.
Here we
exclude dEs with blue cores, since they are known to have different
colors \citep{2008AJ....135..380L}. This RMS-ratio should be one if the scatter is only due to the
    measurement errors and larger than one if there is an intrinsic
    scatter.\footnote{With the binning this is not completely true
    anymore, since flatter relations with the same amount of scatter
    will show a larger RMS-ratio. But if the relation becomes
    significantly flatter, either two relations intersect with an
    offset anyway, or there is a transition region with a flatter CMR;
    in both cases it is different from the relations faint- and
    brightward.} 
In Fig.~1 we show the CMRs along with the RMS-ratio $\rho$.
      
Indeed, the RMS-ratio is enhanced
    between $-20 \textrm{ mag}<M_r<-16 \textrm{ mag}$, indicating an
    intrinsically increased scatter, which could in principle be explained by a transition between two separate relations.
    But the change is steady. Thus also alternative explanations, such as a more varied star formation history with decreasing galaxy mass, seem plausible -- although the decrease in the scatter for the faintest dwarfs would still need to be explained.
 
  One can
    argue about the significance of the RMS-ratio increase for the
    brightest galaxies, since it is just a handful of them
-- nevertheless, this larger scatter might be related to the absence 
of a well-defined CMR at the brightest magnitudes.

\section{Comparison to Semi-Analytical Model}
\label{sec:sam}

The Numerical Galaxy Catalog of \citet{2005ApJ...634...26N} is
based on a high resolution $N$-body simulation in a $\Lambda$CDM universe
\citep{2005PASJ...57..779Y}.  The dark-halo merger trees of the
$N$-body simulation are taken as input for a semi-analytical model
(SAM) of galaxy formation (here a modified version of the Mitaka Model,
\citealt{2004ApJ...610...23N}). 
The SAM models physical processes of galaxy formation and evolution
such as gas radiative cooling, star formation, heating by supernova
explosions (supernova feedback), mergers of galaxies, population
synthesis, and extinction by internal dust and intervening HI
clouds. In particular, the model takes into account the dynamical response
to starburst-induced gas removal after gas-rich mergers (also for cases intermediate
between a purely baryonic cloud and a baryonic cloud fully supported by
surrounding dark matter as in \citealt{1987A&A...188...13Y}). 
 The gravitational well is shallower for the dwarf galaxies and
thus they suffer a more substantial gas loss than giants. This process is not only
important for the galaxy sizes, but also for their metal enrichment histories.

\begin{figure}
\includegraphics[scale=.3,angle=-90]{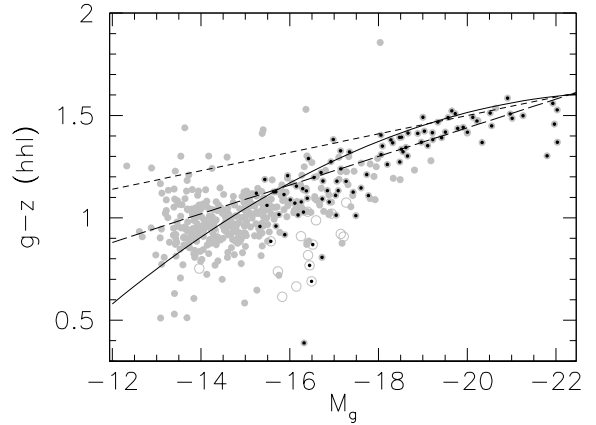}
\caption{Comparison to the data of
    \citet{2006ApJS..164..334F}. Shown are all galaxies in our sample
    with \emph{gray} symbols, again the blue-core dEs with \emph{open
      circles} and the other galaxies with \emph{filled
      circles}. Galaxies common to both samples are indicated by an
    additional  \emph{black dot} in the middle of the other
    symbol. The axes are changed in comparison to Fig.~1 and colors
    are measured in smaller apertures to most closely resemble their
    Fig.~123. The black lines indicate their fits to their data, the
    \emph{short dashed} and \emph{long dashed line} linear fits to a
    brighter ($M_g<-18$) and fainter subsample ($M_g>-18$),
    respectively. The solid line displays the parabola they fitted to
    the whole sample. There is a systematic offset of about 0.1 mag towards bluer colors of the color derived in our study in comparison to \citeauthor{2006ApJS..164..334F}}
\label{fig:ferr}
\end{figure}

In the comparison of the CMR of the Virgo galaxies with the SAM
(Fig.~2), the shapes of the distributions are fairly similar, and the model CMR is indeed not well represented by a
linear relation.
In \citet{2008ApJ...689L..25J} we presented an analogous comparison
for the scaling relation of radius and brightness
and found that the model qualitatively agrees with distribution of the galaxies
in that diagram. Both the CMR and the relation between size and brightness
do not show a linear, nor one common behaviour for dwarfs and giants. 
Nevertheless, in the framework of the SAM both can be explained by the same physical processes, which govern $\Lambda$CDM structure formation, and thus both can be of cosmological origin (see also \citealt{2008arXiv0812.3272C}).

Beside the good 
agreement in the overall shape, 
an offset is observed. This offset could partly be due to
uncertainties of the adopted synthetic stellar population model.\footnote{The offset might here also be influenced to some
  degree by the transformations to Vega magnitudes in the Johnson-Cousins
  filter system. In a diagram of $i-z$ vs. $M_i$ (for which no
  transformations are needed) the observed galaxies lie bluewards of
  the model ones with an smaller absolute offset than in $B-V$
  vs. $M_B$, but the scatter of the observed galaxies is larger due to
  the lower $S/N$ in the $z$-band.}  Furthermore, the relative number of
bright galaxies exceeds the observed one and the luminosity function is clearly different, which could possibly be
  explained with model input physics.

\section{Summary and Discussion}
In the literature, there is no consistent view of commonness or distinctness of 
the color magnitude relation of dwarf and giant early-type galaxies. 
Some studies find a linear relation over the whole range of brightness, 
others find a change of slope. 

We studied the CMR of Virgo cluster early-type
galaxies, based on the wealth of SDSS
imaging data in multiple filter bands. Our main result is that the dwarfs and giants 
do not follow one common \emph{linear} relation. The appearance in the
different colors,  from $u-g$ to $i-z$, is very similar, suggesting that age and
metallicity go hand in hand: the CMR
at shorter wavelengths is more sensitive to changes  in age, while it
is more sensitive to changes in metallicity at longer wavelengths.
Recent studies indeed claim that a combination of the effects of age, metallicity and also $\alpha$-enhancement 
 shape the CMR \citep{2007MNRAS.375..673K,2008ApJ...680.1042L}.

The most direct comparison with other data is possible with
\citet{2006ApJS..164..334F}, who studied early-type galaxies
in the Virgo Cluster with the Hubble Space Telescope's Advanced Camera
for Surveys (ACS) in the ACS Virgo Cluster Survey. In their Fig.~123
they show the CMR in $g-z$, using smaller apertures for the color
measurements. For comparison we repeat the plot and indicate galaxies common to both samples
with black dots in the
symbols (Fig.~3). We also show their fits (to their data): two linear fits
to a brighter and fainter subsample and one parabola for the whole
sample. While a qualitatively good agreement is found, the small offset
to bluer colors of our data is likely explained by a combination of
the effect of different apertures, the atmosphere, and different
physical filters.
While we agree with \citeauthor{2006ApJS..164..334F} that the CMR is
not linear over the whole range, 
 the faint dwarfs in our sample,
which they do not reach (cf.~Fig.~3), do not lie on the extrapolation of their parabolic fit, but rather 
define an overall ``S''-like shape. Furthermore, some of the galaxies
that might drag their CMR to the blue side, at their faint end,
display blue cores from recent star formation activity \citep{2006AJ....132.2432L},
which we excluded.
 
We found an increase of the scatter about the CMR at intermediate
brightness, visible in Fig.~1 through a broadening of the color
distributions within the respective magnitude bins.
This confirms the result that dwarfs and giants do \emph{not} share one common \emph{linear} CMR.
For the brighter and fainter
galaxies the scatter is closer to the photometric errors.

\citet{2005AJ....129...61B} concluded that the CMR is a result of two more fundamental relations: 
the Faber-Jackson relation and a relation between color and velocity
dispersion. Given the slope change of the Faber-Jackson relation from
giants to dwarfs
\citep{2005MNRAS.362..289M,2005A&A...438..491D} a change of slope of
the CMR would actually be expected.

It is important to emphasize that the curved shape of the CMR,
and the possible existence of two adjacent
relations does \emph{not} necessarily imply different formation
scenarios of dwarfs and giants. This is in accordance with previous claims of no distinction between them (\citealt{graham_guzman,2005A&A...430..411G,2006ApJS..164..334F,2008A&A...486..697M}). Instead,
from the good qualitative agreement of the observed CMR shape with a
semi-analytic model, we
conclude that a common origin within the framework of $\Lambda$CDM
structure formation appears plausible for
both giant \emph{and} dwarf  early types. A more detailed
  multiparameter comparison of dwarfs and giants will be
  communicated in a forthcoming paper.

\acknowledgements
We thank Masahiro Nagashima for helpful advice, and the referee for useful comments.
   We are supported within the framework of the Excellence Initiative
    by the German Research Foundation (DFG) through the Heidelberg 
 Graduate School of Fundamental Physics (grant number GSC 129/1).
The study is based on SDSS (http://www.sdss.org/).

\end{document}